\documentclass[twocolumn,showpacs,preprintnumbers]{revtex4}
\input{psfig.sty}
\usepackage{amsmath}
\usepackage{amssymb}
\def\dnu{\Delta\nu}
\def\dT{\Delta T}
\def \QQ{\Theta}
\begin{document}
\draft

\title{
Generalized Haldane Equation and Fluctuation Theorem in the 
Steady State Cycle Kinetics of Single Enzymes
}

\author{Hong Qian}
\address{Department of Applied Mathematics,
University of Washington, Seattle, WA 98195\\
qian@amath.washington.edu}

\author{X. Sunney Xie}
\address{Department of Chemistry and Chemical Biology, 
Harvard University, Cambridge, MA 02138\\
xie@chemistry.harvard.edu}

\date{\today}

\begin{abstract}
Enyzme kinetics are cyclic.  We study a Markov renewal process 
model of single-enzyme turnover in nonequilibrium steady-state 
(NESS) with sustained concentrations for substrates and products.
We show that the forward and backward cycle times have 
idential non-exponential distributions: 
$\QQ_+(t)=\QQ_-(t)$.  This equation generalizes the 
Haldane relation in reversible enzyme kinetics.  In terms 
of the probabilities for the forward ($p_+$) and backward 
($p_-$) cycles, $k_BT\ln(p_+/p_-)$ is shown to be the 
chemical driving force of the NESS, $\Delta\mu$.  More 
interestingly, the moment generating function of the 
stochastic number of substrate cycle $\nu(t)$, 
$\langle e^{-\lambda\nu(t)}\rangle$ follows the fluctuation
theorem in the form of Kurchan-Lebowitz-Spohn-type symmetry.  
When $\lambda$ = $\Delta\mu/k_BT$, we obtain 
the Jarzynski-Hatano-Sasa-type equality: 
$\langle e^{-\nu(t)\Delta\mu/k_BT}\rangle$ $\equiv$ 1 for 
all $t$, where $\nu\Delta\mu$ is the fluctuating chemical work 
done for sustaining the NESS.  This theory suggests possible 
methods to experimentally determine the nonequilibrium driving 
force {\it in situ} from turnover data via single-molecule 
enzymology.
\end{abstract}


\maketitle

\vskip 0.5in

	Most biochemical reactions in a living cell have 
nonzero flux $J$ and nonzero chemical driving force 
$\Delta\mu$.  The nonequilibrium state of such a reaction 
is sustained by continuous material and energy exchange 
with and heat dissipation into its environment \cite{qian05}. 
To understand the state of a biochemical network in an 
open environment, hence, it is necessary to be able to 
experimentally measure both $J$ and $\Delta\mu$ {\it in situ}.  
A large literature exist on measuring $J$, but none exists 
on directly measuring $\Delta\mu$.  One could in 
principle compute $\Delta\mu$ from {\it in situ} measurements 
of the concentrations of the substrate and product of a 
reaction if its equilibrium constant is 
known \cite{qianelson}.  Alternatively, one should be able to 
obtain $\Delta\mu$ from fluctuating cycle kinetics of a single 
enzyme directly.  This possibility has been recently 
investigated in term of stochastic simulations \cite{weimin}. 
Here we exam this idea through a novel analytical model. 

	Enzyme kinetics are complex mainly due to the many 
possible intermediates in the form of enzyme-substrate complexes. 
Recent laboratory measurements with high resolution at the 
single-molecule level give the waiting time distributions
for enzyme cycles \cite{xie}. This motived the present 
Markov renewal process (MRP) model, also known as extended kinetics
model in the theory of motor proteins \cite{KF}.  In terms of 
the MRP, the kinetics of a single enzyme becomes a stochastic 
sequences of forward and backward cycles as function of time.
We shall denote the number of forward and backward cycles by  
$\nu_+(t)$ and $\nu_-(t)$, as shown in Fig. 1.  

\vskip 0.2cm
\begin{figure}[h]
\[
\psfig{figure=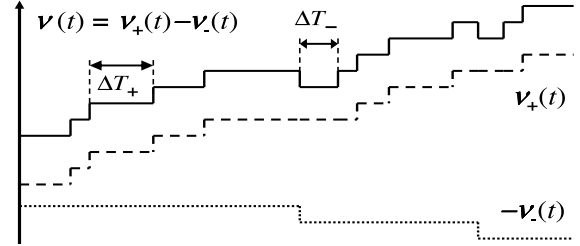,%
width=3.in,height=1.7in,%
bbllx=1.5in,bblly=1.5in,%
bburx=6.5in,bbury=9.in,%
angle=-90}
\]
\caption{The solid line illustrates ideal data on single enzyme 
cycling as function of time, $\nu(t)$, which can be decomposed
into $\nu_+(t)$ and $\nu_-(t)$, shown in dashed and dotted 
lines. The starting positions are arbitary. $\Delta T_+$ 
and $\Delta T_-$ are forward and backward cycle times. 
}
\end{figure}
\vskip 0.3cm

	It is obvious that the cycle time distributions 
give information on the kinetics.  In this letter we show that
the key nonequilibrium thermodynamic quantity, $\Delta\mu$, can 
be obtained from stochastic data on single-enzyme cycle
$\nu(t)\equiv\nu_+(t)-\nu_-(t)$ via two novel equalities
\begin{eqnarray}
  &&\Delta\mu = k_BT\ln\left(\langle\nu_+(t)\rangle/
		\langle\nu_-(t)\rangle\right);
\label{eq1}\\
  &&\langle e^{-\nu(t)\Delta\mu/k_BT}\rangle = 1, \hspace{0.3cm} 
		\forall t,
\label{eq2}
\end{eqnarray}
where $\langle\cdots\rangle$ is ensemble average for repeated 
measurement of $\nu(t)$ in steady-state.  Eq. (\ref{eq1}) 
generalizes a result well known for one-step
chemical reactions \cite{qian05,qian1}.
Eq. (\ref{eq2}) is a version of the fluctuation theorem
(FT) in nonequilibrium statistical mechanics.  The FT for the 
probability distribution of entropy production of a 
nonequilibrium steady-state (NESS) was first discovered
in deterministic dynamical systems \cite{ergodic}.  
Kurchan, Lebowitz and Spohn (KLS) introduced a parallel theory 
in terms of stochastic dynamics \cite{ft} which is more 
appropriate for single enzyme experiments 
\cite{xie,qian2,carlos}.  It was shown that the generating
function, i.e., an exponential average, of a work functional 
$W(t)$ possesses a certain symmetry in the limit of 
$t\rightarrow\infty$.  Crooks introduced a heat functional 
$Q(t)$ and showed the similar symmetry is valid for all 
finite $t$ \cite{crooks}: $c_{\lambda}(t)=c_{1-\lambda}(t)$ where 
$c_{\lambda}(t)=\langle e^{-\lambda Q(t)/k_BT}\rangle$.
Since $Q(t)$ and $W(t)$  differ by a stationary term while both 
increase without bound, Crooks' result immediately yields 
that of KLS's.  The symmetry in the generating function 
implies FT for $Q$ \cite{footnote1}.

	The symmetry implies that 
$\ln\langle e^{-Q(t)/k_BT}\rangle$ = 0.
This is analogous to the Jarzynski equality \cite{jarzy} which 
is surprising since $\langle Q(t)\rangle$ 
$=-\ln e^{-\langle Q(t)\rangle /k_BT}$ 
is the mean heat dissipated from the NESS which 
certainly $\neq$ 0; it should always $>0$. The 
Jarzynski equality provides the possibility to obtain 
a function of state such as free energy from an nonstationary
heat functional $Q(t)$ with finite $t$.  
This was proposed and experimentally tested for mechanical work 
functional on single biological macromolecules such as RNA 
\cite{hs,carlos}.

	The difference between the FTs for $W(t)$ in the limit of
infinite $t$ and for $Q(t)$ with any finite $t$ is crucial to real 
experiments.  In heuristic thermodynamic terms, the work functional 
$W(t)$ \cite{ft} is related to the $\Delta\mu^o$ of a reaction and 
the heat functional $Q(t)$ \cite{crooks}, to the $\Delta\mu$. While 
the former is determined by the transition rate constants, hence 
experimentally accessible in short time, the latter depends on the
stationary probability.  For cyclic enzymatic turnovers, however, 
$W=Q$. Hence, the FT associated with enzyme
cycle kinetics is particularly simple, and 
experimentally accessible \cite{weimin}. 
Generalizing the Jarzynski equality to open-system, 
Hatano and Sasa's equality for NESS \cite{jarzy} also suggested 
a possibility of computing chemical driving force for 
single-molecule chemical reactions in NESS,  
see \cite{weimin,seifert}.  

	To show Eqs. (\ref{eq1}) and (\ref{eq2}), there
are two strategies. One is based on traditional Markov models,
i.e., master equations, for single enzyme kinetics. Then 
both equations can be show as conseqeunces of the existing 
FTs \cite{ft,crooks}. An alternative, more insightful approach
however, is to model the kinetics in terms of a MRP with
cycle kinetics.  In this new model, we shall show a suprizing
equality between the forward and backward cycle time 
distributions: $\Theta_+(\tau)=\Theta_-(\tau)$.  With this equality,
Eq. (\ref{eq1}) becomes obvious, and Eq. (\ref{eq2}) can be
shown in elementary terms, in the Eqs. (\ref{klseq1})-(\ref{equal1}) 
below. 

	The equality $\QQ_+(\tau)=\QQ_-(\tau)$ turns out to 
be a very important, unknown relation in enzyme kinetics. This is
a key result of this work.  It has to do with microscopic 
reversibility. There are experimental edvidence for it, as well 
as theoretical models proving equal mean time 
$\langle\Delta T_+\rangle =\langle\Delta T_-\rangle$ 
\cite{yanagida,Kolomeisky}.  We shall give a proof for the equal 
distribution with sequential enzyme kinetics. The proof for
more general systems will be published elsewhere \cite{wangqian}. 

	{\em The Markov renewal process (MRP) model.---}
Detailed kinetic scheme of an enzyme catalyzed biochemical 
reaction $A\rightleftharpoons B$ is usually 
very complex \cite{segel}.  But if one concerns only with the net 
number of steady-state turnovers from $A$ to $B$, $\nu(t)$,
it can be represented by a continuous-time, 
discrete-state one dimensional random walk with cumulative 
cycle time distribution functions $\QQ_{\pm}(t)$ for the 
forward and the backward stochastic transition times
$\Delta T_+$ and $\Delta T_-$: 
$\QQ_{\pm}(0)=0$, $\QQ_{\pm}(\infty)=1$, and $\QQ_{\pm}(t)$ 
are nondecreasing.  This is a class of
stochastic models known as MRP \cite{mrp} which
has wide applications in single-enzyme kinetics and motor
protein stepping \cite{mrpqian,KF}.  See Fig. 2 in which 
$w_{\pm}(t)=p_{\pm}\QQ_{\pm}(t)$, and $p_++p_-=1$.
$p_+$ ($p_-$) is the eventual probability of the enzyme 
binding $A$ ($B$) and converting it to $B$ ($A$).
We shall also denote $w(t)$ = $w_+(t)+w_-(t)$.
   
	We discover that a necessary condition for the 
Eqs. (\ref{eq1}) and (\ref{eq2}) is the cycle-time 
distributions for the forward and backward steps being equal: 
$\QQ_+(t)=\QQ_-(t)$.  We call this equality generalized Haldane 
equation \cite{footnote2}.

\begin{figure}[h]
\[
\psfig{figure=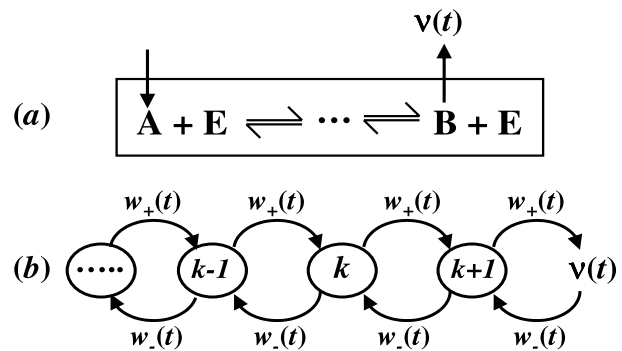,%
width=3.3in,height=2.in,%
bbllx=1.5in,bblly=1in,%
bburx=6.9in,bbury=9.5in,%
angle=-90}
\]
\caption{(a) A schematics for an enzyme reaction converting
substrate $A$ to product $B$. In a NESS, the concentrations for 
$A$ and $B$, $c_A$ and $c_B$, are controlled through feedback 
by an experimenter.  The cumulative number of $B$ taken out 
by the time $t$ is denoted by $\nu(t)$, 
$-\infty<\nu(t)<\infty$.  (b) The integer valued $\nu(t)$ is
most naturally modeled by a random walk with forward and 
backward time distributions $w_+(t)$ and $w_-(t)$ \cite{KF}. 
}
\end{figure}
\vskip 0.3cm

	The position of the random walker in Fig. 2b, $\nu(t)$, 
models the net number of enzyme turnover.  Let 
$\nu_0=0,\dnu_1, \dnu_2,...,\dnu_{\ell},...$ ($\dnu=\pm 1$) be the 
successive increments of turnover number, and 
$T_0=0,\dT_1,\dT_2,...,\dT_{\ell},...$ ($\dT\ge 0$) be the 
corresponding increments in time.  Then the probabilistic 
meaning of $w_{\pm}(t)$ is the joint probability for 
continuous $\Delta T$ and binary $\Delta\nu$:
\begin{equation}
          w_{\pm}(t) = \Pr\{\dnu_{\ell}=\pm 1,\dT_{\ell}\le t\}, 
			\quad (\ell\ge 1).
\end{equation}
The equation $\QQ_+(t)=\QQ_-(t)$ leads to $w_{\pm}(t)=p_{\pm}w(t)$.
That is the random variables $\Delta\nu_{\ell}$ and $\Delta T_{\ell}$
are statistically independent!

 {\it $\QQ_+(t)=\QQ_-(t)$ and microscopic reversibility.---} 
To show this equality for forward and backward cycles, we 
consider a sequential enzyme reaction as shown in Fig. 3a and 
a corresponding exit problem \cite{TK} shown in 3b. Starting 
at the central position $E$, $w_+(t)$ and $w_-(t)$ 
are the cumulative probabilities of reaching $B+E$ and $A+E$. 
Since only the first and last steps are irreversible, 
$w_+(t)$ and $w_-(t)$ both have $2n+1$ exponential
terms with same eigenvalues, one of which is $0$. 
Both thus can be written as
$a_0 + a_1e^{-\lambda_1t} + a_2e^{-\lambda_2t} 
... + a_{2n}e^{-\lambda_{2n}t}$.
With some straightforward algebra, it can be shown that
for all $ 0\le m\le 2n$ \cite{footnote5}, 
\begin{equation}
        \frac{1}{w_1w_2...w_n}\frac{d^mw_-(0)}{dt^m}
	= \frac{1}{u_1u_2...u_n}\frac{d^mw_+(0)}{dt^m}.
\end{equation}
Since functions $w_+(t)$ and $w_-(t)$ are completely
determined by these initial conditions which satisfy the 
linear algebraic system, we have 
\begin{equation}
            \frac{w_+(t)}{w_-(t)} \equiv \prod_{\ell=1}^n
           \left(\frac{w_{\ell}}{u_{\ell}}\right)
	 = e^{-\Delta\mu/k_BT},
\label{wratio}
\end{equation}
independent of $t$.  That is $\QQ_-(t)=\QQ_+(t)$. 

	The meaning of the equality now becomes clear: 
We recall that $u_1$ and $w_n$ are pseudo-first order
rate constants: $u_1=u_1^oc_A$ and $w_n=w_n^oc_B$. 
In a chemical equilibrium:
\begin{equation}
          \frac{c_B}{c_A} = \frac{u_1^ou_2...u_{n-1}u_n}
		{w_1w_2...w_{n-1}w_n^o},
\label{eqcond}
\end{equation}
that is $w_+(t)=w_-(t)$.   Therefore in a chemical 
equilibrium, not only on the average $w_+(\infty)=w_-(\infty)$, 
i.e., the forward flux equals the backward flux, but 
the detailed kinetics for the transition time distributions
has to be equivalent: {\em There is absolutely no statistical
difference between the forward and backward reactions.}
In a NESS when (\ref{eqcond}) is not held true, 
$w_+(t)\neq w_-(t)$.  But the difference is only in
the total probability $p_+=w_+(\infty)$ and $p_-=w_-(\infty)$, 
the distribution functions $\QQ_+(t)=\QQ_-(t)$ still hold true.  
This equality is essencial to the KLS symmetry below.
It is known that microscopic reversibility has to 
be satisfied even when a mesoscopic system is in a 
nonequilibrium steady-state \cite{ft}.

\begin{figure}[h]
\[\psfig{figure=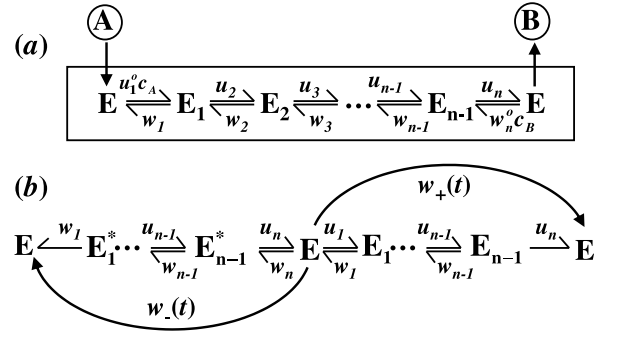,%
width=3.4in,height=2.in,%
bbllx=1.in,bblly=0in,%
bburx=7.5in,bbury=11.0in,%
angle=-90}
\]
\caption{(a) A schematic for an enzyme reaction 
converting $A$ to $B$. 
The transition time distribution of a single enzyme 
converting an $A$ to $B$, $w_+(t)$, and converting a
$B$ to $A$, $w_-(t)$, are intimately related to the
exit problem shown in (b) in which  $u_1$ and $w_n$ are 
pseudo-first order rate constants which depend
on the concentrations of $A$ and $B$, respectively:
$u_1=u_1^oc_A$, $w_n=w_n^oc_B$.  The scheme in (b) has
been used to compute steady-state one-way flux in 
T.L. Hill's theory on biochemical cycle kinetics 
\cite{hill,qian1}.
}
\end{figure}

	{\em Kurchan-Lebowitz-Spohn (KLS) symmetry and FT.---}
For $k$ successive number of renewal events (forward 
plus backward turnovers) within time $[0,t]$, let us denote 
$(\nu_k,T_k)$ 
$=\sum_{\ell=1}^k (\Delta\nu_{\ell},\Delta T_{\ell})$.  
The moment generating function for $\nu(t)$ is
\begin{eqnarray}
      g_{\lambda}(t) &\equiv& 
	\big\langle e^{-\lambda\nu(t)}\big\rangle
\label{klseq1}\\
	&=& \sum_{n=-\infty}^{\infty} e^{-\lambda n}
 \sum_{k=0}^{\infty}\Pr\{\nu_k=n,T_k\le t, T_{k+1}>t\}
\nonumber
\end{eqnarray}
\begin{eqnarray} 
	&=&\sum_{k=0}^{\infty}\left(\sum_{n=-k}^k
	e^{-\lambda n}\Pr\{\nu_k=n\}\right)
	\Pr\{T_k\le t,T_{k+1}>t\}
\label{sep}\\
	&=&\sum_{k=0}^{\infty}\left(p_+e^{-\lambda}+
		p_-e^{\lambda}\right)^k
	\Pr\{T_k\le t,T_{k+1}>t\}
\label{glambda}
\end{eqnarray}
The Eq. (\ref{sep}) is obtained due to the independence
between $\nu_k$ and $T_k$.  Then from Eq. (\ref{glambda})
we have the KLS symmetry
\begin{equation}
	g_{\lambda}(t) = g_{\lambda^*-\lambda}(t),
	\hspace{0.2cm} \forall t,
\label{gcs} 
\end{equation}
where $\lambda^*=\ln (p_+/p_-)$.  Furthermore, 
\begin{equation}
	g_{\lambda^*}(t) \equiv \big\langle 
		e^{-\nu(t)\Delta\mu/k_BT} \big\rangle 
	= g_0(t) = 1,
\label{equal1}
\end{equation}
if $\ln(p_+/p_-)=\Delta\mu/k_BT$ holds true. 
We recognize that $\nu(t)\Delta\mu$ is the external 
chemical work done to the system in NESS. Hence  
Eq. (\ref{equal1}) is analogous to the Jarzynski 
equality for a cycle.

	{\it Chemical driving force for NESS enzyme 
cycle.---} If we let the $t\rightarrow\infty$ in Eq.
(\ref{wratio}), we have 
$\ln(p_+/p_-)\equiv\lambda^*=\Delta\mu/k_BT$, which is 
needed in deriving Eq. (\ref{equal1}). This 
generalizes the well-known result for single step chemical 
reactions \cite{hill,qian1} to any complex enzyme
reaction cycle. 

	We are now also in a position to show 
Eq. (\ref{eq1}). The mean number of net turnovers can 
be computed from the $g_{\lambda}(t)$ given in
Eq. (\ref{glambda}):
\begin{eqnarray}
      \langle\nu(t)\rangle &=& \langle\nu_+(t)\rangle
	- \langle\nu_-(t)\rangle = -\left[
	\frac{dg_{\lambda}(t)}{d\lambda}\right]_{\lambda=0}
\\
    &=& (p_+-p_-)\sum_{k=0}^{\infty}k
		\Pr\{T_k\le t,T_{k+1}>t\}
\\
   &=& (p_+-p_-)\times\textrm{mean \# of cycles in time $t$}
\end{eqnarray}
Therefore,  
$\frac{\langle\nu_+\rangle}{\langle\nu_-\rangle}=\frac{p_+}{p_-}$.
Furthermore, in the limit of large $t$ \cite{TK}, 
$\langle\nu(t)\rangle$ $\approx$$(p_+-p_-)t/\langle T_1\rangle$,
where $\langle T_1\rangle$ $=$ $\int_0^{\infty} tdw(t)$
is the mean time for one cycle, forward or
backward.  When $p_+ = p_-$, the steady-state flux
$J=$ $\lim_{t\rightarrow\infty}\langle\nu(t)\rangle/t$
$=0$ as expected.  When $p_+>p_-$, $J>0$.

	{\it Summary.---} Studying enzyme-catalyzed 
biochemical reactions {\it in situ} requires methods for 
measuring $\Delta\mu$, the NESS chemical driving force.  
Currently none exists.  We propose obtaining $\Delta\mu$
from stochastic cycle data of a single enzyme molecule, 
$\nu(t)$, via (i) an equality similar to that of Jarzynski 
and Hatano-Sasa: 
$\langle e^{-\nu(t)\Delta\mu/k_BT}\rangle$ = 1; or simply (ii)
$k_BT\ln (\langle\nu_+(t)\rangle/ \langle\nu_-(t)\rangle)$.  
We developed a MRP model for the enzyme cycles with arbitary
complex mechanism, and have found an equality between the forward
and backward cycle time distributions based on microscopic 
reversibility.  This equality is a generalization of the what
is known as Haldane relation for the reversible enzyme kinetics,
and a recent results in \cite{Kolomeisky}.  
The model enables us to establish a FT and above equalities (i) 
and (ii) for any $t$. Noting that $(1/t)\langle \nu(t)\rangle$ = $J$, 
one thus obtains both the flux $J$ and the driving force 
$\Delta\mu$ for a reaction in NESS from the fluctuating 
$\nu(t)$.   The statistical accuracies associated
with these measurements were discussed in \cite{weimin}. 

	We thank C. Asbury, M. Fisher, C.  Jarzynski, 
A. Kolomeisky, W. Min, U. Seifert, Y.-Y. Shi, A. Szabo, 
H.-Y. Wang, and J. Yu for discussions, and anonymous 
reviewers for very helpful comments.


\small

\begin{references}

\bibitem{qian05}
For recent reviews, see 
H. Qian, J. Phys. Cond. Matt. {\bf 17}, S3783-S3794 (2005); 
H. Qian and D.A. Beard, Biophys. Chem. {\bf 114}, 213-220 (2005). 

\bibitem{qianelson}
H. Qian and E.L. Elson, Proc. Natl. Acad. Sci. USA, {\bf 101}, 
2828-2833 (2004); W.J. Heuett and H. Qian, J. Chem. Phys. 
{\bf 124}, 044110 (2006). 

\bibitem{weimin}
W. Min, L. Jiang, J. Yu, S.C. Kou, H. Qian, and X.S. Xie,
Nano Lett. {\bf 5}, 2373-2378 (2005).

\bibitem{xie}
B.P. English, et al. Nature Chem. Biol. {\bf 2}, 87-94 (2006);
S.C. Kou, B.J. Cherayil, W. Min, B.P. English
and X.S. Xie, J. Phys. Chem. B {\bf 109}, 19068-19081 (2005);
X.S. Xie and H.P. Lu, J. Biol. Chem.  {\bf 274}, 15967-15970 
(1999). 
 

\bibitem{KF}
A.B. Kolomeisky and M.E. Fisher, J. Chem. Phys. {\bf 113},
10867-10877 (2000). 

\bibitem{qian1}
H. Qian, Phys. Rev. E. {\bf 64}, 022101 (2001).

\bibitem{ergodic}
G. Gallavotti, Phys. Rev. Lett. {\bf 77}, 4334 (1996);
G. Gallavotti and E.G.D. Cohen, {\it ibid}. {\bf 74},
2694 (1995);
J.R. Dorfman, {\it An Introduction to Chaos in Nonequilibrium
Statistical Mechanics} (Cambridge Univ. Press, New York, 1999); 
D.-Q. Jiang, M. Qian and M.-P. Qian, {\it Mathematical Theory of 
Nonequilibrium Steady-state}, LNM Vol. 1833 
(Springer, New York, 2004).

\bibitem{ft}
J. Kurchan, J. Phys. A. {\bf 31}, 3719-3729 (1998);
J.L. Lebowitz and H. Spohn, J. Stat. Phys. {\bf 95},
333-365 (1999); P. Gaspard, J. Chem. Phys. {\bf 102}, 
8898-8905 (2004). 

\bibitem{qian2}
H. Qian, Phys. Rev. E. {\bf 65}, 016102 (2002);
H. Qian, J. Phys. Chem. B. {\bf 106}, 2065-2073 (2002).

\bibitem{carlos}
J. Liphardt, S. Dumont, S.B. Smith, I. Tinoco, and
C. Bustamante, Science, {\bf 296}, 1832-1835 (2002);
E.H. Trepagnier, C. Jarzynski, F. Ritort, G.E. Crooks, 
C.J. Bustamante, and J. Liphardt, {\it Proc. Natl.
Acad. Sci. USA}, {\bf 101}, 15038-15041 (2004). 

\bibitem{crooks}
G. Crooks, Phys. Rev. E. {\bf 60}, 2721-2726 (1999);
D.-Q. Jiang, M. Qian, and F.-X. Zhang, J. Math. Phys.  
{\bf 44}, 4174-4188 (2003); 
U. Seifert, Phys. Rev. Lett. {\bf 95}, 040602 (2005);
H. Qian, J. Phys. Chem. B {\bf 109}, 23624-23628 (2005). 

\bibitem{footnote1}
Symmetry $c_{\lambda}=c_{1-\lambda}$ means 
$\sum_a e^{-\lambda a}\Pr\{Q/k_BT=a\}=
\sum_a e^{-\lambda a}e^{-a}{\Pr\{Q/k_BT=-a\}}$
$\forall$ $\lambda$. Hence 
$\frac{\Pr\{Q/k_BT=a\}}{\Pr\{Q/k_BT=-a\}}$ $=e^a$.
The symmetry is a mathematical version of the FT.

\bibitem{jarzy}
C. Jarzynski, Phys. Rev. Lett.  {\bf 78}, 2690-2693 (1997).
While Jarzynski's equality was first established for 
time-dependent conservative systems, it was latter shown 
that it is also valid for NESS of open systems, see 
T. Hatano and S.-i. Sasa, Phys. Rev. Lett. {\bf 86}, 3463-3466
(2001).

\bibitem{hs}
G. Hummer and A. Szabo, Proc. Natl. Acad. Sci. USA {\bf 98},
3658-3661 (2001). 

\bibitem{seifert}
U. Seifert, Europhys. Lett. {\bf 70}, 36-41 (2005).  

\bibitem{yanagida}
M, Nishiyama, H, Higuchi and T. Yanagida, 
Nature Cell Biol. {\bf 4}, 790-797 (2002);
N.J. Carter and R.A. Cross, Nature, {\bf 435} 308-312 (2005).

\bibitem{Kolomeisky}
A.B. Kolomeisky, E.B. Stukalin, and A.A. Popov, 
Phys. Rev. E. {\bf 71}, 031902 (2005). 

\bibitem{wangqian}
H. Wang and H. Qian, preprint. 

\bibitem{segel}
I.H. Segel, {\it Enzyme Kinetics} (John-Wiley Interscience, New
York, 1975).

\bibitem{mrp}
E. Cinlar, {\it Introduction to Stochastic Processes}
(Prentice-Hall, Englewood Cliffs, 1975). A MRP is a combination 
of Markov processes with renewal theory. It has non-exponential 
sojourn time for each transition.  It becomes a Markov process 
if the distributions of the sojourn times are all exponential 
and independent of the next state; it becomes a Markov chain 
if the sojourn times are all equal to one, and it becomes a
renewal process if there is only one state.

\bibitem{mrpqian}
H. Qian, Biophys. Chem. {\bf 67}, 263-267 (1997);
H. Qian and E.L. Elson, {\it idib}. {\bf 101}, 565-576 
(2002).

\bibitem{footnote2}
The Haldane relation in reversible enzyme kinetics \cite{segel} 
states that the steady-state velocities for forward and backward 
enzyme reactions, $V_f$ and $V_r$, have same saturation dependence 
on the substrate and product concentrations: 
$V_f=$ $\frac{\alpha_f[S]}{1+[S]/K_S+[P]/K_P}$, 
$V_r=$ $\frac{\alpha_r[P]}{1+[S]/K_S+[P]/K_P}$, where
$\frac{\alpha_f}{\alpha_r}=K_{eq}$, the equilibrium constant
between $S$ and $P$.  Also see \cite{Kolomeisky}.

\bibitem{TK}
H.M. Taylor and S. Karlin, {\it An Introduction to Stochastic
Modeling}, 3rd Ed. (Academic Press, New York, 1998).

\bibitem{footnote5}
The initial condition leads to 
$\sum_{\ell=0}^{2n} a_{\ell}=$
$\sum_{\ell=1}^{2n} a_{\ell}\lambda_{\ell}=...$
$\sum_{\ell=1}^{2n} a_{\ell}\lambda_{\ell}^{n-1}=0$,
but
$d^nw_-(0)/dt^n=$
$\sum_{\ell=1}^{2n}a_{\ell}\lambda_{\ell}^n=$
$w_1w_2...w_n$ and
$d^nw_+(0)/dt^n=$
$\sum_{\ell=1}^{2n}a_{\ell}\lambda_{\ell}^n=$ 
$u_1u_2...u_n$. Furthermore,
$d^{n+1}w_-(0)/dt^{n+1}=$
$\sum_{\ell=1}^{2n}a_{\ell}\lambda_{\ell}^{n+1}=$
$-w_1...w_n\sum_{k=1}^n(w_k+u_k)$,
$d^{n+1}w_+(0)/dt^{n+1}=$
$\sum_{\ell=1}^{2n}a_{\ell}\lambda_{\ell}^{n+1}=$
$-u_1...u_n\sum_{k=1}^n(w_k+u_k)$.



\bibitem{hill}
T.L. Hill, {\it Free Energy Transduction and Biochemical
Cycle Kinetics} (Springer-Verlag, New York, 1989);
{\it Free Energy Transduction in Biology} (Academic 
Press, New York, 1977).


\end{references}
\end{document}